\documentclass{jpsj2}
\def\runauthor{Yuki \textsc{Fuseya},
Hiroshi \textsc{Kohno} and Kazumasa \textsc{Miyake}}

\title{Renormalization Group Technique Applied
to the Pairing Interaction of the Quasi-One-Dimensional Superconductivity}

\author{Yuki \textsc{FUSEYA}
\thanks{Present address: Department of Physics, Graduate School of Science, Nagoya University,
Nagoya 464-8602, Japan.}
\thanks{E-mail address: fuseya@slab.phys.nagoya-u.ac.jp}, 
Hiroshi \textsc{Kohno}
and Kazumasa \textsc{Miyake}
 }

\inst{Department of Physical Science, Graduate School of 
Engineering Science, Osaka University,
Toyonaka, Osaka 560-8531, Japan}

\newcommand{\simg}{\stackrel{>}{_\sim}}
\newcommand{\siml}{\stackrel{<}{_\sim}}
\newcommand{\Tc}{T_{\rm c}}
\newcommand{\ima}{{\rm i}}

\abst{
	A mechanism of the quasi-one-dimensional (q1d) superconductivity is 
	investigated by applying the renormalization group techniques 
	to the pairing interaction.
	With the obtained renormalized pairing interaction,
	the transition temperature $\Tc$ and corresponding gap function
	are calculated by solving the linearized gap equation.
	For reasonable sets of parameters,
	$\Tc$ of $p$-wave triplet pairing is higher than that of $d$-wave singlet pairing
	due to the one-dimensionality of interaction.
	These results can qualitatively explain the superconducting properties of 
	q1d organic conductor (TMTSF)$_2$PF$_6$
	and the ladder compound Sr$_2$Ca$_{12}$Cu$_{24}$O$_{41}$.
}

\kword{quasi-one-dimensional superconductor, renormalization group, (TMTSF)$_2$PF$_6$,
Sr$_2$Ca$_{12}$Cu$_{24}$O$_{41}$}

\begin{document}
\maketitle

\section{Introduction}

	The nature of the attractive interaction and the type of Cooper pairing in quasi-one-dimensional
	(q1d) superconductivity have led to lively discussions in these decades.
	The firstly discovered and well-known q1d superconductor is the Bechgaard salts (TMTSF)$_2$PF$_6$
	\cite{Jerome}.
	Its superconducting (SC) phase is in proximity to the spin density wave (SDW) phase,\cite{TMTSFphase}
	so that it has been believed so far that the $d$-wave singlet pairing
	is realized due to the antiferromagnetic spin fluctuations.
	\cite{MSV,Scalapino,Emery}
	The recent experiments, however, are apparently against this expectation.
	The unsaturated behavior of the upper critical field $H_{\textrm{c}2}$
	and the absence of the Knight shift suppression
	below the transition temperature on (TMTSF)$_2$PF$_6$
	strongly suggests the triplet pairing\cite{Hc2,Hc22,Knight,Knight2}.
	The $H_{\textrm{c}2}$ and the Knight shift data on the ladder material
	Sr$_2$Ca$_{12}$Cu$_{24}$O$_{41}$ also suggest the triplet pairing\cite{Fujiwara}.

	In order to investigate the mechanism of the q1d superconductivity,
	various theoretical attempts have been done.
	Previous studies on the q1d Hubbard model have
	concluded the most favorable pairing symmetry is the $d$-wave singlet,
	which is reasonable in view of the unconventional superconductivity
	near the antiferromagnetic or SDW phase.\cite{Shimahara, Kino}
	On the other hand, some recent studies pointed out that triplet pairing
	can be realized assisted by charge fluctuations.
	\cite{KurokiFLEX,Fuseya,Tanaka}
	In a study based on third-order perturbation theory, which do not rely on a particular type of fluctuation 
	but treat the moderate effect of electron-electron interaction,
	it was concluded that the $d$-wave singlet pairing is much more stable 
	than the triplet one.
	\cite{Nomura}
	The quantum Monte Carlo study concluded that the $f$-wave triplet pairing 
	can compete with (but cannot prevail over) 
	the $d$-wave singlet one\cite{KurokiQMC}.
	However, the validity of the existing theoretical methods are not so obvious in q1d systems
	for the following reasons.
	\begin{enumerate}
		\item[(a)] Perturbative corrections are not small at low energy and low temperature
		where logarithmic singularities appear in each order of perturbation series.
		Therefore the higher-order corrections should also be taken into account.
		
		\item[(b)] Not only the specific fluctuations, such as spin or charge fluctuations,
		are developed, but also the vertex corrections have singularities.
		Therefore we should take into account each singular diagram on an equal footing.
	\end{enumerate}
	These are the effects of the one-dimensionality.

	The reason (a) suggests the low-order (third or even fourth order) perturbation theory
	is not enough,
	and (b) suggests that the random phase approximation (RPA) type approximation 
	(including fluctuation exchange approximation),
	which emphasizes some particular fluctuations, such as spin fluctuations\cite{FLEXnote},
	might lead to incorrect conclusion.

	The renormalization group (RG) techniques
	have been applied to one-dimensional (1d) systems
	as a powerful method which can overcome these difficulties.\cite{Solyom}
	In the conventional RG method for 1d system (so-called g-ology) 
	only two types of superconductivity,
	$s$-wave singlet and $p$-wave triplet, which are fully gapped on the Fermi surface,
	have been considered, and the superconductivity with gap having nodes
	on the Fermi surface have been out of considerations.
	It is, however, expected from the previous theoretical studies 
	that there is a possibility that the gap has nodes on the Fermi surface in q1d and even in purely 1d systems\cite{Fuseya}.

	The purpose of this paper is to investigate the mechanism of the q1d superconductivity
	by the newly developed RG method
	and to discuss the SC mechanism in repulsively interacting q1d electron systems.
	This method can treat superconductivity with nodes,
	and can take into account the effects of the one-dimensionality.

	In \S 2 we present our theoretical method.
	The obtained gap functions and the transition temperatures for various situations
	are shown in \S 3.
	We give an interpretation of the results,
	and discuss the mechanism of the q1d superconductivity in \S 4.
	Section 5 is devoted to conclusions.
	We also present detailed momentum-dependence of the pairing interaction
	and a phenomenological discussion of a possibility
	of an odd-energy superconductivity in appendix A and B, respectively.

\section{Theory}

\subsection{Outline}
	The q1d systems with a tiny interchain hopping ($t_{\perp}$), i.e., nearly 1d systems,
	are well-described by the 1d RG.
	Suzumura and Fukuyama \cite{Suzumura} showed the $t_{\perp}$-dependence
	of the transition temperature, which exhibit the transition temperature
	of the SDW, $T_{\rm SDW}$, increases as $T_{\rm SDW} \propto t_{\perp}^{\theta}$
	for small $t_{\perp}$ and has a maximum at a certain $t_{\perp}^{*}$,
	then turns down.
	(Here $\theta$ is a positive constant of the order of unity.)
	The superconductivity is realized above $t_{\perp}^*$,
	but the 1d RG cannot be applied for such relatively large $t_{\perp}$.
	So, we need a theoretical method which is appropriate for the q1d system
	above $t_{\perp}^*$, where the effects of the one-dimensionality 
	is expected to be still remain.
	This is just what we shall present in this paper.

	The outline of the present method is as follows.
	First, we apply the RG technique to the irreducible part of the vertex, namely, the pairing interaction.
	With this renormalized pairing interaction, 
	we solve the linearized gap equation 
	\begin{eqnarray}
	\Delta (\textbf{k})=-\sum_{\textbf{k}'}V(\textbf{k},\textbf{k}')\frac{\Delta (\textbf{k}')}{2\xi_{\textbf{k}'}}
	\tanh \frac{\xi_{\textbf{k}'}}{2\Tc},
	\label{gapeq}
	\end{eqnarray}
	in the weak-coupling formalism 
	to determine the transition temperature $\Tc$
	and the corresponding gap function $\Delta (\textbf{k})$ 
	without any assumption on the functional form of the gap.

	Solving the linearized gap equation 
	is equivalent to studying the Cooper instability with the sum of all matrix elements which are 
	irreducible in the particle-particle channel in the sense that  
	they cannot diagrammatically be divided into two parts by cutting two parallel electron lines.
	The pairing interaction $V(\textbf{k},\textbf{k}')$ thus has to be irreducible with respect to 
	the particle-particle channel,
	otherwise the particle-particle channel is multiply counted when we solve the gap equation.
	We need to calculate not the full but the irreducible vertex,
	which cannot be obtained by the conventional RG method.
%

	\subsection{Multiplicative RG technique}

	The procedure of the multiplicative RG method is as follows.
	The multiplicative property of the Green function, vertex function and other
	physical quantities (e.g., response functions) allows one to improve
   perturbative results by 
	solving the RG Lie equation.
	Assume that a quantity $A(\omega /E_0, g_i )$ has a scaling property and satisfies
	\begin{eqnarray}
		A\Biggl( \frac{\omega}{E_0 '}, g_i ' \Biggr)
		=z_A \Biggl( \frac{E_0 '}{E_0}, g_i \Biggr)
		A\Biggl( \frac{\omega}{E_0 }, g_i  \Biggr),
 	\label{scalingeq}
	\end{eqnarray}
	then $A(\omega /E_0, g_i )$ obeys the differential equation (Lie equation)
	\begin{eqnarray}
		& &\frac{\partial }{\partial x} \ln A(x, g_{1 ||}, g_{1 \perp}, g_2 ) \nonumber \\
		&&=\frac{1}{x}\frac{\partial}{\partial \xi}
		\biggl\{ \ln A \biggl( \xi , g_{1 ||}'(x, g_{1 ||}, g_{1 \perp}, g_2 ),
		g_{1 \perp}'(x, g_{1 ||}, g_{1 \perp}, g_2 ),
		g_{2}'(x, g_{1 ||}, g_{1 \perp}, g_2 ) \biggr) \biggr\} _{\xi = 1}, \nonumber \\
		\, \label{Lieeq}
	\end{eqnarray}
	where $x=\omega /E_0 $, $T/E_0$ or $v_F k/E_0 $, and the prime sign indicates the renormalized 
	quantities.
	Notations for coupling constants here follows the conventional g-ology \cite{Solyom}.
	The scattering process with $g_1$ corresponds to backward scattering, whose momentum transfer
   is of the order of $2k_F$. 
   The processes with $g_2$ are forward scattering.
   The couplings with parallel (opposite) spins are indicated by the subscript $||$ ($\perp $).

	The first step of the RG method is to calculate
	the Green function and vertices perturbatively.
	Using these perturbative forms in the right-hand side of the Lie equation
	and solving it,
	we obtain the renormalized solution for the invariant couplings.
	Once the renormalized invariant couplings are determined,
	we can also obtain other quantities (e.g., response functions) by solving 
	the respective Lie equations.

	In the present RG method, the first step, where the renormalized invariant coupling is obtained,
	follows from the conventional g-ology\cite{Solyom}.
	The essential and specific point of our RG method is the second step,
	where we do not calculate the renormalized response functions, 
	but the renormalized pairing interactions (which is two-particle irreducible in the particle-particle channel).

\subsection{Irreducible vertices}
	\begin{figure}[tbp]
	\begin{center}\leavevmode
	\includegraphics[width=8cm]{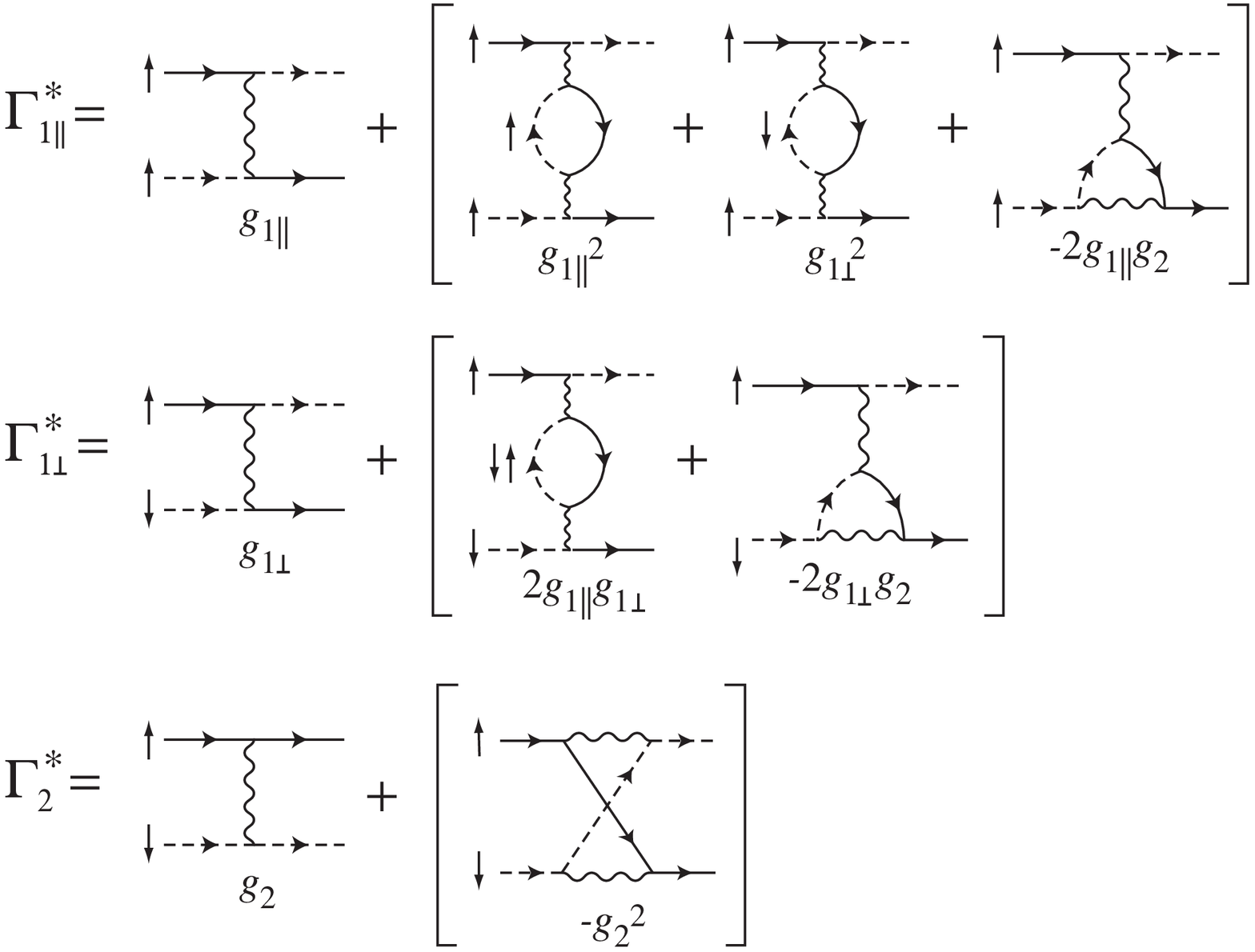}
	\caption{Diagrams of the irreducible vertices $\Gamma_{1||, 1\perp , 2}^*$
	up to $g_{1||, 1\perp , 2}^2$.
	The diagrams enclosed by [\, ] have contributions 
	$(2\pi v_F )^{-1} \ln (\omega /E_0 )$.
	The imaginary parts are neglected.
	}\label{irrgall}\end{center}
	\end{figure}

	A relation between 
	the irreducible part of the four-point vertex $\Gamma^*$ 
	and the full vertex is given by the equation
	\begin{eqnarray}
	\Gamma_i (\omega ) &=& \Gamma_i^* (\omega )
	+{\rm i}\!\!\int \frac{dp}{2\pi}\frac{d\omega '}{2\pi} 
	\Gamma_i^* (k_F , -k_F, p, -p; 3\omega /2,-\omega /2, \omega ', \omega -\omega ')
	\nonumber \\
	& &\times G(p, \omega ')G(-p, \omega -\omega ')
	\Gamma_i (p, -p, -k_F , k_F ;\omega ', \omega -\omega ', \omega /2, \omega /2),
	\nonumber \\
	\,
	\end{eqnarray}
	where $\Gamma$'s are the full vertices and $G$ is the full Green function.
	Suffices of $\Gamma$'s have usual meanings\cite{Solyom}.
	The irreducible parts are given, up to the lowest-order corrections, 
	i.e., $\mathcal{O} (g_i^2 )$, in the form
	\begin{eqnarray}
	\Gamma_{1||}^* (\omega)&=& g_{1||} + \frac{1}{2\pi v_F}
	(g_{1||}^2 +g_{1\perp}^2-2g_{1||}g_2 )
	\Biggl(
	\ln \frac{\omega}{E_0}-\frac{\pi \ima}{2}
	\Biggr), \\
	\Gamma_{1\perp}^* (\omega )&=& g_{1\perp} + \frac{1}{\pi v_F}
	(g_{1||}g_{1\perp}-g_{1\perp}g_2)
	\Biggl(
	\ln \frac{\omega}{E_0}-\frac{\pi \ima}{2}
	\Biggr), \\
	\Gamma_2^* (\omega ) &=& g_2 + \frac{1}{2\pi v_F}
	(-g_2^2 )
	\Biggl(
	\ln \frac{\omega}{E_0}-\frac{\pi \ima}{2}
	\Biggr) .
	\end{eqnarray}
	The corresponding diagrams are shown in Fig. \ref{irrgall}.
	In order to consider not only these second-order contributions,
	but also the higher-order ones,
	we employ the RG technique.

	Introducing the dimensionless irreducible vertices $\widetilde{\Gamma}_i^*$ as
	$\Gamma_i^* \equiv g_i \widetilde{\Gamma}_i^*$,
	the scaling differential equations (cf. eq. (\ref{Lieeq}))
	are obtained (up to one-loop order\cite{1loop}) as
	\begin{eqnarray}
	\frac{\partial}{\partial x}\ln \widetilde{\Gamma}_{1||}^* (x)
	&=&\frac{1}{x}\frac{1}{2\pi v_F}\Biggl\{
	g_{1||}'(x)+\frac{g_{1\perp}^{'2}(x)}{g_{1||}(x)}-2g_2'(x)
	\Biggl\} , \label{irredifeq1}\\
	\frac{\partial}{\partial x}\ln \widetilde{\Gamma}_{1\perp}^* (x)
	&=&\frac{1}{x}\frac{1}{\pi v_F}
	\bigl\{
	g_{1||}'(x) - g_2' (x)
	\bigl\} , \label{irredifeq2}\\
	\frac{\partial}{\partial x}\ln \widetilde{\Gamma}_2^* (x)
	&=&-\frac{1}{x}\frac{1}{2\pi v_F}
	g_2'(x) ,\label{irredifeq3}
	\end{eqnarray}
	which are different from the scaling equation for the full vertices $\Gamma _i $.\cite{Solyom}
	These scaling equations (\ref{irredifeq1})-(\ref{irredifeq3})
	can be solved in a closed form in the case $g_{1||}=g_{1\perp}$ 
	with use of the g-ology results, $g_1 '(x)=g_1 /[1-(g_1 /\pi v_F ) \ln x]$
	and $g_2 '(x) = g_2 -g_1 /2 + g_1 '(x)/2$, as
	\begin{eqnarray}
	\widetilde{\Gamma}_1^* (x) &=& \bigl[ 1-(g_1 /\pi v_F )\ln x \bigr] ^{-1/2}
	x^{-\alpha}, \label{loggam1}\\
	\widetilde{\Gamma}_2^* (x) &=&\bigl[ 1-(g_1 /\pi v_F )\ln x \bigr] ^{1/4}
	x^{-\alpha/2}, \label{loggam2}
	\end{eqnarray}
	where $\alpha \equiv (g_2 - g_1 /2 )/\pi v_F $.
	Near the fixed point, i.e., in the low-energy region, 
	the logarithmic correction is less dominant than the power term,
	and is considered to be weakened when we apply higher order RG (e.g. two-loop RG).
	The renormalized irreducible vertex in the low-energy region is thus given by
	\begin{eqnarray}
	\widetilde{\Gamma}_1^* &\simeq &(\omega /E_0 )^{-\alpha}, \label{renpair1}\\
	\widetilde{\Gamma}_2^* &\simeq &(\omega /E_0 )^{-\alpha/2}. \label{renpair2}
	\end{eqnarray}
	(The effect of the logarithmic correction will be discussed in \S 4.)
	Parenthetically, we note that the full vertices $\Gamma_{1, 2}$
	are renormalized as $\Gamma_1 (\omega ) \to 0, \Gamma_2 (\omega ) \to g_2 -g_1 /2 $ (constant)
	\cite{SolyomI}.
	This is due to the fact that the zero-sound type and Cooper type singularity compensate each other.
	By contrast, the renormalized irreducible vertex $\Gamma _{1, 2}^*$ have the power-law singularity.
	This is understood as the lack of the compensation.
	(The diagrams which are effectively contained in the renormalized $\Gamma_i^*$ are
	displayed in Fig. \ref{Hubbard}.)

	Using these renormalized irreducible vertices,
	we can construct the pairing interaction both for singlet and 
	triplet channel.

\subsection{Pairing interactions for quasi-one-dimensional systems}

	The pairing interactions 
	are given by
	\begin{eqnarray}
	V_{||}^{\rm t}(k+k')&=& -g_{1||}(k+k') +g_{2||}(k+k'), \label{V1dgt1}\\
	V_{\perp}^{\rm t}(k+k')&=& -g_{1\perp}(k+k') +g_{2\perp}(k+k'), \label{V1dgt2}
	\end{eqnarray}
	for triplet, and
	\begin{eqnarray}
	V^{\rm s}(k+k')&=& g_{1\perp}(k+k') +g_{2\perp}(k+k'), \label{V1dgs}
	\end{eqnarray}
	for singlet.
	(See Appendix \ref{Vtrisin} for details.)
	Using the results of the renormalization applied to the irreducible vertices,
	(\ref{renpair1}) and (\ref{renpair2}),
	the pairing interaction for $g_{1||}=g_{1\perp}$ is given by
	\begin{eqnarray}
	V^{\rm t}(k+k') &=& -g_1 \Biggl( \frac{k+k'-2k_F}{2k_0} 
	\Biggr) ^{-\alpha } +g_2 \Biggl( \frac{k+k'-2k_F}{2k_0} 
	\Biggr) ^{-\alpha /2}, \hspace{0.5cm}\\
	V^{\rm s}(k+k') &=& g_1 \Biggl( \frac{k+k'-2k_F}{2k_0} 
	\Biggr) ^{-\alpha } +g_2 \Biggl( \frac{k+k'-2k_F}{2k_0} 
	\Biggr) ^{-\alpha /2}.
	\end{eqnarray}

\begin{figure}[tb]
	\begin{center}\leavevmode
	\includegraphics[width=6cm]{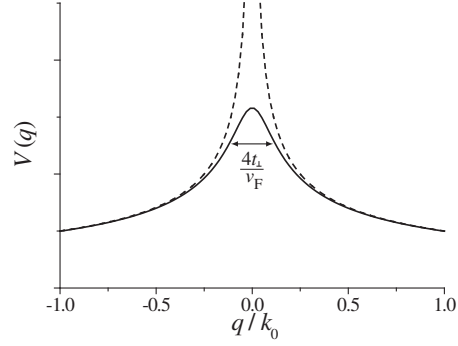}
	\caption{Model pairing interaction in q1d system.
	The solid line is the q1d interaction suppressed by the interchain hopping $t_{\perp}$.
	The dashed line indicates the 1d interaction which has a power-law singularity at $q=0$.
	}\label{suppress}\end{center}
	\end{figure}

	With this pairing interaction, we can determine the SC transition
	temperature $\Tc$ and the gap function without any assumptions on the form of the gap functions
	by solving the linearized gap equation (\ref{gapeq}) numerically.
	Here, we take the q1d dispersion
	\begin{eqnarray}
	\xi ({\bf k}) = -2t \cos k_x -2t_{\perp}\cos k_y -\mu , \label{q1ddispersion}
	\end{eqnarray}
	where $t$ ($t_{\perp}$) is the intrachain (interchain) transfer integral,
	and $\mu$ is the chemical potential.
	Of course, when we solve the gap equation (\ref{gapeq}) in q1d systems,
	we have to take a q1d pairing interaction, i.e., $V({\bf k, k'})$
	must have dependence on $k_x , k_y , k_x' $ and $ k_y '$.
	It is, however, naturally expected that
	the dominant character of the pairing interaction
	is determined by the 1d pairing interaction.
	Following this idea, we take the quasi-one dimensionality into account
	as follows.
	For a q1d dispersion $\xi (k_x , k_y)=v_F (|k_x |-k_F )-2t_{\perp}\cos k_y -\mu$,
	the logarithmic singularity in the Cooper and the zero-sound channels are cutoff as
	\begin{eqnarray}
	\pm \frac{1}{2\pi v_F}\ln \Biggl( 
	\frac{2t_{\perp}}{E_0} \Biggr),
	\end{eqnarray}
	so that the q1d Cooper and zero-sound channel may have a form
	\begin{eqnarray}
	\pm \frac{1}{2\pi v_F}\ln \Biggl( 
	\frac{\sqrt{(q/2)^2 + (t_{\perp}/v_F )^2 }}{k_0} \Biggr),
	\end{eqnarray}
	where $q=|k+k' -2k_F |$.
	Namely, the interchain hopping suppresses
	the singularity of the interaction (see Fig. \ref{suppress}).
	The renormalized irreducible vertices are also suppressed,
	and then we obtain the q1d pairing interaction as follows:
	\begin{eqnarray}
	V_{\rm t}^{\rm q1d}(q, t_{\perp}) &=& -g_1 
	\Biggl( \frac{\sqrt{(q/2)^2 + (t_{\perp}/v_F )^2 }}{k_0}
	\Biggr) ^{-\alpha } 
	+g_2 \Biggl( \frac{\sqrt{(q/2)^2 + (t_{\perp}/v_F )^2 }}{k_0}
	\Biggr) ^{-\alpha /2},\label{q1dtri} \nonumber \\
	\, \\
	V_{\rm s}^{\rm q1d}(q, t_{\perp}) &=& 
	g_1 \Biggl( \frac{\sqrt{(q/2)^2 + (t_{\perp}/v_F )^2 }}{k_0}
	\Biggr) ^{-\alpha } 
	+g_2 \Biggl( \frac{\sqrt{(q/2)^2 + (t_{\perp}/v_F )^2 }}{k_0}
	\Biggr) ^{-\alpha /2}. \label{q1dsin} \nonumber \\
	\,
	\end{eqnarray}
	Although the pairing interaction (\ref{q1dtri}) and (\ref{q1dsin})
	depends only on the momentum parallel to the chain,
	the quasi-one-dimensionality is taken into account.
	
	Using the pairing interaction, (\ref{q1dtri}) and (\ref{q1dsin}),
	and the q1d dispersion (\ref{q1ddispersion}) in (\ref{gapeq}), we calculate
	the transition temperature $\Tc$ and the gap function at $\Tc$.
	We focus on the case of 1/4 filling as a model of (TMTSF)$_2$X compounds,
	and take $k_F = \pi /4 $ ($k_0 = \pi /4$).
	We take finer mesh near the Fermi surface, i.e.,
	1000 points in the region $0<k_x <\pi $ and
	the minimum size of the mesh is $\Delta k_{x}^{\rm min}=6.0 \times 10^{-8}$ 
	(the energy resolution is $1.7 \times 10^{-7}t \sim 0.5$mK).

	In the present framework, the gap equation gives finite $\Tc$ even for $t_{\perp}=0$.
	This is due to the mean field approximation for the SC instability, namely,
	the present method fully takes into account the zero-sound type diagrams,
	but partly does the Cooper type diagrams.
	For $t_{\perp}\sim 0$, there is an interference between them generally,
	which leads $\Tc$ to zero.\cite{Suzumura}
	The region in which the interference is remarkable is $T>t_{\perp}$.
	Therefore the valid region of the present theory is $T<t_{\perp}\ll t$.
	This region seems to be narrow, but the unknown and interesting phenomena happen there.
	The behavior for $t_{\perp}<T\ll t$ would smoothly continue to the g-ology with finite $t_{\perp}$.
	\cite{Suzumura}

\section{Results}
	
	\begin{figure}[tb]
	\begin{center}\leavevmode
	\includegraphics[width=9cm]{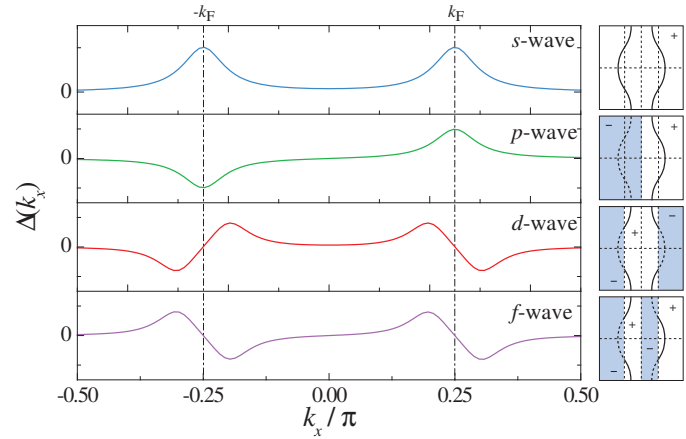}
	\caption{Variety of the superconductive gap symmetry in q1d systems. (left panels)
	The vertical line at $k=k_F = \pi /4$ indicates the Fermi point in purely 1d case ($t_{\perp}=0$).
	The right panels show the Fermi surface, where the solid (dashed) line indicates
	$\Delta (k_x )>0$ ($<0$).
	}\label{symmetry}\end{center}
	\end{figure}
	The gap function of q1d systems are roughly classified as shown in Fig. \ref{symmetry}.
	It is true that, in q1d systems, 
	we cannot classify the gap symmetry by the spherical-wave basis such as $s, p, d$.
	We use such terminology to express how many times the gap changes sign
	on the Fermi surface, i.e., ``$s$-wave" does not change its sign,
	``$p$-wave" changes twice, ``$d$-wave" four times, and so on.
%
%
%
%
	Although the gap functions depend only on $k_x$ since
	the pairing interaction $V_{\bf k, k'}^{\rm s, t}$ does,
	the interchain hopping $t_{\perp}$ slightly warps the Fermi surface, and
	the $d$- and $f$-wave gap thus have 4 line nodes on the Fermi surface.
	In other words, it seems sufficient for these four types of pairing symmetry
	to consider only $k_x$-dependence of the pairing interaction.

	\begin{figure}
	\begin{center}\leavevmode
	\includegraphics[width=8cm]{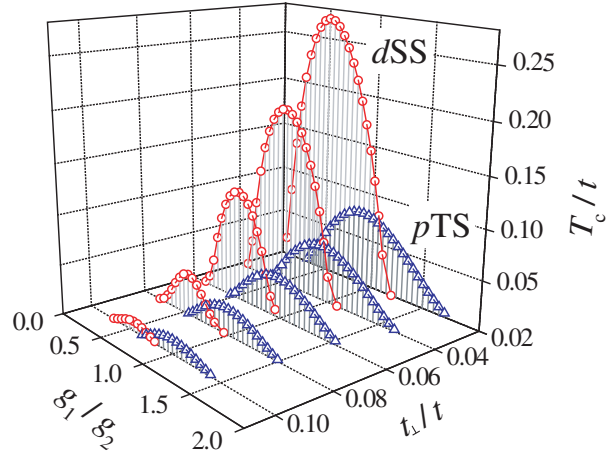}
	\caption{Interchain hopping $t_{\perp}/t$ - backward scattering $g_1 /g_2$ - transition temperature $\Tc$
	phase diagram for $g_2 = 0.8 \pi v_F$.
	The open circles indicate $\Tc$ of $d$-wave singlet superconductivity ($d$SS),
	the open triangles indicate that of $p$-wave triplet superconductivity ($p$TS).
	}\label{3Dphase}\end{center}
	\end{figure}
	\begin{figure}
	\begin{center}\leavevmode
	\includegraphics[width=8cm]{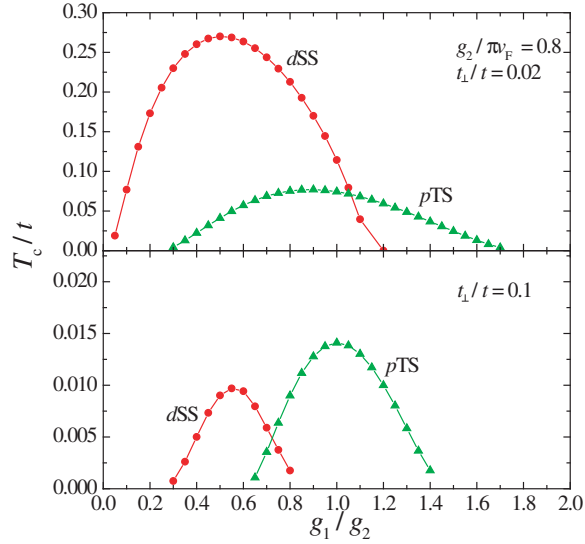}
	\caption{Superconducting transition temperatures as functions of $g_1 / g_2$ for $t_{\perp} /t = 0.02$ (top)
	and for $t_{\perp}/t=0.1$ (bottom).
	The circles indicate $\Tc$ of $d$-wave singlet superconductivity ($d$SS),
	the triangles indicate that of $p$-wave triplet superconductivity ($p$TS).
	}\label{ty}\end{center}
	\end{figure}
	\begin{figure}[tbh]
	\begin{center}\leavevmode
	\includegraphics[width=8cm]{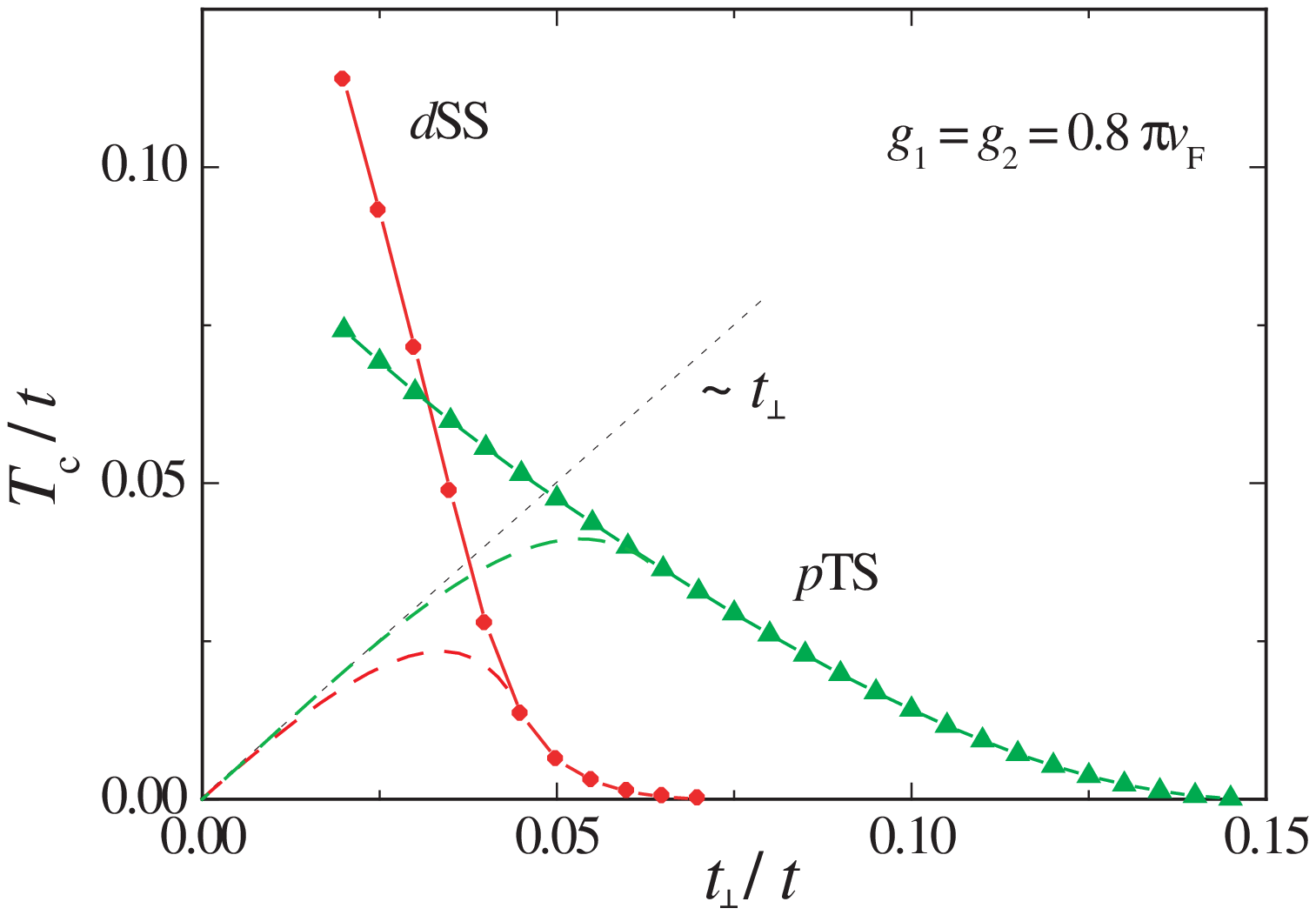}
	\caption{Interchain hopping $t_{\perp}$ dependence of $\Tc$'s in the case of 
	$g_1 = g_2 = 0.8\pi v_F $, the Hubbard model case.
	The dashed lines indicate the suppression due to the 1d fluctuation.
	}\label{U08}\end{center}
	\end{figure}
	The results of calculation are shown in Fig. \ref{3Dphase}
	for various values of $t_{\perp}$ and $g_1$ and for fixed $g_2 = 0.8 \pi v_F$.
	The region we are interested in is the SDW (CDW) phase in the context of g-ology,
	i.e., $0\le g_1 /g_2 \le 2$\cite{Solyom}. 
	In this parameter region, we obtain two types of pairing symmetries,
	$p$-wave triplet ($p$TS) and $d$-wave singlet ($d$SS).
	For $t_{\perp} =0.02t$, $d$SS is dominant for $0\le g_1 /g_2 \le 1.0$
	and $p$TS is for $1.1 \le g_1 /g_2 \le 1.7$. (Fig. \ref{ty})
	In the case of $g_1 = g_2$, which corresponds to the Hubbard model,
	$\Tc$ of $d$SS is higher than that of $p$TS.
	When we increase $t_{\perp}$, however, $\Tc$ of $d$SS is suppressed more
	drastically than that of $p$TS, so that $p$TS is the most stable for
	$t_{\perp}\simg0.03t$ as shown in Fig. \ref{U08}.
	Figure \ref{3Dphase} and Fig. \ref{U08} tell us that $d$SS is 
	``weaker" against $t_{\perp}$ than $p$TS, i.e., against the suppression of the singularity
	in the pairing interaction.

	In the present theoretical framework, the 1d fluctuation is considered in the pairing interaction,
	but the SC transition is determined within the mean field approximation.
	When we consider the 1d fluctuation effect upon the transition,
	it is expected that the $\Tc$'s are suppressed as $\Tc \sim t_{\perp}$\cite{Suzumura}
	as schematically shown by the dashed lines in Fig. \ref{U08}.
	The SDW phase would overhang for $t_{\perp}/t\siml 0.1$\cite{Yamaji}.
	Considering the long-range nature of the Coulomb interaction,
	we may take $g_1 \siml g_2$, $t_{\perp}/t \sim 0.1$
	for the observed SC state in (TMTSF)$_2$PF$_6$.
	According to our calculation for $t_{\perp} /t =0.1$,
	the $p$TS arises for $g_1 /g_2 \simg 0.7$.
	Even in such case, the field-induced $p$TS is realized by applying Zeeman magnetic field,
	while the ground state is $d$SS for $g_1 / g_2 \siml 0.7$,
	since the singlet pairing is suppressed due to the paramagnetic effect while
	the triplet pairing is not.
	The signs of the triplet pairing in (TMTSF)$_2$PF$_6$ are mainly observed under the magnetic field.
	\cite{Hc2,Hc22,Knight,Knight2}
	So, there is a possibility that the field-induced $p$TS is realized in (TMTSF)$_2$PF$_6$.
	The present results are also consistent with the experiments of Sr$_2$Ca$_{12}$Cu$_{24}$O$_{41}$,
	which shows coherence peak at $\Tc $ and exponential decrease below $\Tc $ in $1/T_1$ 
	suggesting the gap.\cite{Fujiwara}
	($p$TS is fully gapped for the present open Fermi surface.)

\section{Discussion}
	
	Our result that the $p$TS instability is rather strong for parameters
	appropriate to the Hubbard model is different from other 
	theoretical studies as described in \S 1.
	In order to clarify the difference between the present study and the 
	previous ones,
	let us concentrate on the Hubbard model case ($g_1 = g_2 \equiv U$) for the moment.

\begin{figure}
	\begin{center}\leavevmode
	\includegraphics[width=8cm]{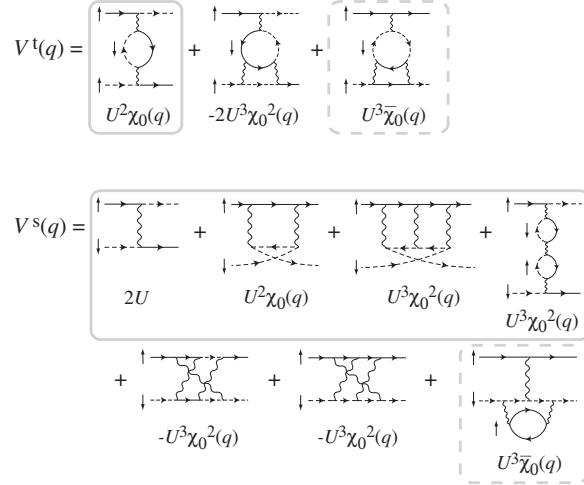}
	\caption{Diagrammatical expression of the effective pairing interaction 
	for the Hubbard model on the basis
	of the second-order logarithmic perturbation.
	The bare spin susceptibility $\chi_{0}(q)$ corresponds to the zero-sound channel,
	i.e., $\chi_0 (q) = -(1/2\pi v_F )\ln |(q-2k_F )/2k_0 |$. ($\overline{\chi}_0 (q)\equiv \chi_0 (q)/2\pi v_F $)
	Diagrams enclosed with gray-solid line are
	included in the RPA-type approximation,
	and that with gray-dashed line are not included 
	in the present one-loop approximation.
	}\label{Hubbard}\end{center}
	\end{figure}

	The diagrams of the pairing interaction for the Hubbard model
	on the basis of the perturbation expansion up to the quadratic order in logarithmic singularity
	are displayed in Fig. \ref{Hubbard}.
	Roughly speaking, negative $V^{\rm t}$ promotes the $p$-wave triplet pairing,
	and positive $V^{\rm t}$ promotes the $f$-wave.
	Likewise negative $V^{\rm s}$ promotes the $s$-wave singlet pairing,
	and positive $V^{\rm s}$ promotes the $d$-wave. (See Appendix \ref{AppB}.)
	On the basis of RPA-like approximation, which consider the diagrams
	enclosed by gray-solid lines in Fig. \ref{Hubbard},
	we have
	\begin{eqnarray}
	V^{\rm t, RPA} (q) &=& U^2 \chi_0 (q) >0, \\
	V^{\rm s, RPA} (q) &=& 2U + U^2 \chi_0 (q) 
	+2U^3 \chi_0^2 (q) >0,
	\end{eqnarray}
	where $q=k+k'$
	and $\chi_0 (q) \equiv -(1/2\pi v_F )\ln |(q-2k_F )/2k_0 |$ 
	is the zero-sound or particle-hole channel,
	which corresponds to the bare spin susceptibility.\cite{note}
	These form indicate that $f$-wave ($d$-wave) is stable for triplet (singlet) pairing,	
	and the $d$SS is more favorable than $f$TS
	due to the large $U^3 \chi_0^2 (q)$ term,
	which is consistent with the previous studies based on the antiferromagnetic
	spin fluctuation theory \cite{Shimahara,Kino,KurokiFLEX}.
	When we consider vertex corrections as well, we have
	\begin{eqnarray}
	V^{\rm t} (q) &=& U^2 \chi_0 (q) -2U^3 \chi_0^2 (q) 
	+U^3 \overline{\chi}_0 (q) <0, \label{Vt3}\\
	V^{\rm s} (q) &=& 2U + U^2 \chi_0 (q) + U^3 \overline{\chi}_0(q) >0,
	\end{eqnarray}
	where $\overline{\chi}_0(q)\equiv \chi_0 (q)/2\pi v_F$.
	Since the bare spin susceptibility $\chi_0 $ diverges at $q=2k_F$, 
	even if $U$ is small,
	the relation $V^{\rm t}<0$ holds near $q\sim 2k_F$.
	For the singlet pairing, $V^{\rm s} (q)$ is still positive, but is smaller than $V^{\rm s, RPA} (q)$,
	because the vertex corrections cancel out the large positive $U^3 \chi_0^2 (q)$ term.
	Therefore, $\Tc$ of the $d$-wave pairing is drastically suppressed by the vertex corrections,
	which is consistent with the third-order perturbation study \cite{Nomura}.
	In this sense, the vertex corrections are important for the triplet pairing to prevail over the singlet pairing.
	The vertex corrections, moreover, can shift from $f$-wave to $p$-wave within 
	the triplet pairing.
	In the case of the triplet pairing, $V^{\rm t} (q)$ becomes negative, so that
	the pairing symmetry changes from $f$-wave to $p$-wave due to the most singular
	$-2\widetilde{U}^3 \chi_0^2 (q)$ term.
	Even if we consider the suppression of $\chi_0 (q)$ due to $t_{\perp}$ or temperature,
	$\chi_0^2 (q)$ will remain enhanced compared to $\chi_0 (q)$ in q1d systems, i.e.,
	the first and the third terms in eq. (\ref{Vt3}) cannot cancel out the second term,
	so that $p$-wave triplet pairing is considered to be robust\cite{sasaki}.
	This consideration based on the second order logarithmic (third order in $U$) perturbation
	does not tell us that the vertex correction of $\mathcal{O}(U^3)$ makes an essential contribution,
	but only tells us the singular vertex correction has a potential to change the pairing symmetry.
	This is the very 1d effect.
	In order to clarify its influence, however, we have to examine 
	higher-order term of the singular vertex corrections.
	The validity of the simple perturbation theory becomes more questionable due to the logarithmic 
	singularity when the one-dimensionality is strong.
	The present results are obtained by considering both the vertex corrections and 
	the higher-order logarithmic singularities within the one-loop RG.
%
%
%
	The one-loop approximation is sufficient for the case $g_{1||}\ge |g_{1\perp}|$, and thus
	the present method is appropriate to evaluate the higher-order leading terms.
	(Note that our calculation may overestimate this $\chi_0^2 (q)$ term since
	one-loop approximation does not contain the next-leading diagrams
	enclosed by gray-dashed line in Fig. \ref{Hubbard}.)

%
%
	
	Incidentally, the logarithmic term in eq. (\ref{loggam1}) and (\ref{loggam2})
	may not be negligible within the one-loop RG.
	In q1d systems, the power-law singularity is suppressed as 
	$A(q)\sim (q^2 /4+ t^2_{\perp}/v_F^2 )^{-\alpha /2}$,
	which is comparable to the logarithmic singularity
	near $q\sim 2t_{\perp}/v_F $.
	If these logarithmic corrections are relevant,
	it suppresses the backward scattering $\Gamma_1^*$
	and enhances the forward scattering $\Gamma_2^*$,
	namely, $\Tc$ of $p$TS is suppressed and that of $d$SS is enhanced.
	Though it is, 
	this suppression (enhancement) for $p$TS ($d$SS) is considered to be small
	since the logarithmic corrections are generally irrelevant when we apply 
	the higher-loop approximation.


\section{Conclusion}
	
	We applied the multiplicative renormalization group method
	to the irreducible vertex as the pairing interaction and calculated the transition temperatures
	and the gap functions.
	The renormalized pairing interactions have the power-law singularity
	as $\Gamma_1^* \propto (q/2k_0 )^{-\alpha}$, $\Gamma_2^* \propto (q/2k_0 )^{-\alpha /2}$,
	here $\alpha = (g_2  -g_1 /2)/\pi v_F$.
	Comparing their exponents, we may conclude that the irreducible vertex for backward scattering
	is much singular than the forward one.
	The solution of the gap equation shows that the $p$-wave triplet pairing can be realized
	in q1d system ($t_{\perp}/t \sim 0.1 $, $g_1 / g_2 \siml 1.0$), prevailing over the $d$-wave singlet pairing.
	When we take into account the logarithmic correction to the irreducible vertices, 
	$\Tc$ of the $p$-wave triplet state can be suppressed and
	that of the $d$-wave singlet state is enhanced.
	This logarithmic correction might be an artifact
	of the one-loop approximation,
	so we need further investigation whether this logarithmic correction
	is irrelevant or not.
	The two-loop approximation of the irreducible vertex, however,
	does not obey the scaling assumption, and thus we must take another route
	to settle this issue.
	Nonetheless, our new method allows us to recognize the one-dimensional effect
	which enhances the $p$-wave triplet instability in q1d systems,
	which has not been noticed in the previous theoretical works.
	The obtained results may give interpretations of the spin-triplet behavior of q1d materials
	(TMTSF)$_2$PF$_6$ and Sr$_2$Ca$_{12}$Cu$_{24}$O$_{41}$.

\section*{Acknowledgement}

One of the author (Y. F. ) is benefited from the conversation with Y. Suzumura,
T. Giamarchi, and N. Dupuis.
This work was supported by
the 21st Century COE Program and a Grant-in-Aid for
Creative Scientific Research from Japan Society for the
Promotion of Science (JSPS).
Y.F. is supported by
Research Fellowships of JSPS for Young Scientists,
and a Grant-in-Aid for Scientific Research on Priority Areas of Molecular Conductors 
(No. 15073103) from the Ministry of Education, Culture, Sports, Science and Technology, Japan.

\appendix

\section{Pairing interactions for singlet and triplet pairing}\label{Vtrisin}

Here we mention the momentum dependence of the interaction.
	The first correction to the irreducible vertex is the zero-sound type diagram,
	which give contributions
	\begin{eqnarray}
	\chi_0 (k\pm k')&=& \textrm{i}\!\!\int \!\!\frac{dp}{2\pi}\frac{d\omega}{2\pi}G_+ (p, \omega )G_- (p-k\pm k') 
	\nonumber \\
	&=&-\frac{1}{2\pi v_F}\ln \frac{v_F |k\pm k' -2k_F |}{E_0}.
	\end{eqnarray}
	
	\begin{figure}[tb]
	\begin{center}\leavevmode
	\includegraphics[width=7cm]{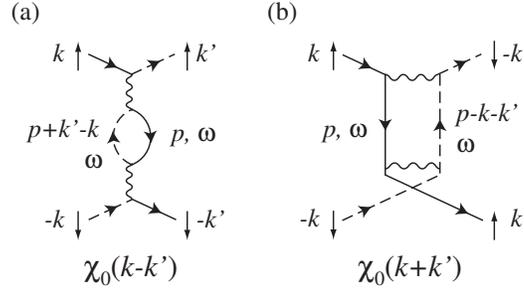}
	\caption{Two different diagrams of zero-sound type. 
	(a) zero-sound channel contributing to the backward scattering;
	(b) that to the forward scattering.
	}\label{zero-sound diagram}\end{center}
	\end{figure}
	As seen in Fig. \ref{zero-sound diagram},
	the zero-sound type contribution to the attractive interaction in the BCS
	Hamiltonian can be classified into two groups
	\begin{eqnarray}
	\chi_0 (k-k') b_{k'\sigma }^{\dagger}a_{-k'\sigma '}^{\dagger}
	b_{-k\sigma '}a_{k\sigma}, \label{chig1}\\
	\chi_0 (k+k') a_{k'\sigma }^{\dagger}b_{-k'\sigma '}^{\dagger}
	b_{-k\sigma '}a_{k\sigma} \label{chig2}
	\end{eqnarray}
	Equation (\ref{chig1}), displayed in Fig. \ref{zero-sound diagram} (a),
	corresponds to the backward scattering,
	and eq. (\ref{chig2}), displayed in Fig. \ref{zero-sound diagram} (b),
	to the forward scattering.
	Therefore, we obtain the pairing interaction of the present 1d model as follows:
	\begin{eqnarray}
	\mathcal{H}_{\rm int}&=&\frac{1}{L}\sum_{k, k'}
	\bigl\{
	g_{1||}(k-k')\bigl[ b_{k'\uparrow}^{\dagger}a_{-k'\uparrow}^{\dagger}
	b_{-k\uparrow}a_{k\uparrow}
	+b_{k'\downarrow}^{\dagger}a_{-k'\downarrow}^{\dagger}
	b_{-k\downarrow}a_{k\downarrow}\bigr]  \nonumber \\
	&+ &g_{1\perp}(k-k')\bigl[ b_{k'\uparrow}^{\dagger}a_{-k'\downarrow}^{\dagger}
	b_{-k\downarrow}a_{k\uparrow}
	+b_{k'\downarrow}^{\dagger}a_{-k'\uparrow}^{\dagger}
	b_{-k\uparrow}a_{k\downarrow}\bigr]  \nonumber \\
	&+& g_{2 ||}(k+k')\bigl[ a_{k'\uparrow}^{\dagger}b_{-k'\uparrow}^{\dagger}
	b_{-k\uparrow}a_{k\uparrow}
	+a_{k'\downarrow}^{\dagger}b_{-k'\downarrow}^{\dagger}
	b_{-k\downarrow}a_{k\downarrow}\bigr]  \nonumber \\
	&+& g_{2 \perp}(k+k')\bigl[ a_{k'\uparrow}^{\dagger}b_{-k'\downarrow}^{\dagger}
	b_{-k\downarrow}a_{k\uparrow}
	+a_{k'\downarrow}^{\dagger}b_{-k'\uparrow}^{\dagger}
	b_{-k\uparrow}a_{k\downarrow}\bigr] 
	\bigr\} \nonumber \\
	&=&\frac{1}{L}\sum_{k, k'}
	\bigl\{
	-g_{1||}(k+k')\bigl[ a_{k'\uparrow}^{\dagger}b_{-k'\uparrow}^{\dagger}
	b_{-k\uparrow}a_{k\uparrow}
	+a_{k'\downarrow}^{\dagger}b_{-k'\downarrow}^{\dagger}
	b_{-k\downarrow}a_{k\downarrow}\bigr]  \nonumber \\
	&- &g_{1\perp}(k+k')\bigl[ a_{k'\downarrow}^{\dagger}b_{-k'\uparrow}^{\dagger}
	b_{-k\downarrow}a_{k\uparrow}
	+a_{k'\uparrow}^{\dagger}b_{-k'\downarrow}^{\dagger}
	b_{-k\uparrow}a_{k\downarrow}\bigr]  \nonumber \\
	&+& g_{2 ||}(k+k')\bigl[ a_{k'\uparrow}^{\dagger}b_{-k'\uparrow}^{\dagger}
	b_{-k\uparrow}a_{k\uparrow}
	+a_{k'\downarrow}^{\dagger}b_{-k'\downarrow}^{\dagger}
	b_{-k\downarrow}a_{k\downarrow}\bigr]  \nonumber \\
	&+& g_{2 \perp}(k+k')\bigl[ a_{k'\uparrow}^{\dagger}b_{-k'\downarrow}^{\dagger}
	b_{-k\downarrow}a_{k\uparrow}
	+a_{k'\downarrow}^{\dagger}b_{-k'\uparrow}^{\dagger}
	b_{-k\uparrow}a_{k\downarrow}\bigr] 
	\bigr\} \nonumber \\
	&=&\frac{1}{L}\sum_{k, k'}
	\bigl\{ -g_{1||}(k+k')+g_{2||} (k+k') \bigr\} \nonumber \\
	& &\times \bigl[ a_{k'\uparrow}^{\dagger}b_{-k'\uparrow}^{\dagger}
	b_{-k\uparrow}a_{k\uparrow}
	+a_{k'\downarrow}^{\dagger}b_{-k'\downarrow}^{\dagger}
	b_{-k\downarrow}a_{k\downarrow} \bigr] 
	\nonumber \\
	&- &g_{1\perp}(k+k')
	\bigl[ a_{k'\downarrow}^{\dagger}b_{-k'\uparrow}^{\dagger}
	b_{-k\downarrow}a_{k\uparrow}
	+a_{k'\uparrow}^{\dagger}b_{-k'\downarrow}^{\dagger}
	b_{-k\uparrow}a_{k\downarrow}\bigr] 
	\nonumber \\
	&+& g_{2\perp}(k+k')
	\bigl[a_{k'\uparrow}^{\dagger}b_{-k'\downarrow}^{\dagger}
	b_{-k\downarrow}a_{k\uparrow}
	+a_{k'\downarrow}^{\dagger}b_{-k'\uparrow}^{\dagger}
	b_{-k\uparrow}a_{k\downarrow}\bigr] \bigr\} .
	\end{eqnarray}
	Here we put $k' \to -k'$ in the second equality.
	The pairing interaction can be separated into spin-symmetric 
	and spin-antisymmetric parts as
	\begin{eqnarray}
	\mathcal{H}_{\rm int}&=&\frac{1}{L}\sum_{k, k'}
	\bigl\{ -g_{1||}(k+k')+g_{2||}(k+k') \bigr\} \nonumber \\
	&\times &\bigl[ a_{k'\uparrow}^{\dagger}b_{-k'\uparrow}^{\dagger}
	b_{-k\uparrow}a_{k\uparrow}
	+a_{k'\downarrow}^{\dagger}b_{-k'\downarrow}^{\dagger}
	b_{-k\downarrow}a_{k\downarrow} \bigr] \nonumber \\
	&+ &\bigl\{ -g_{1\perp}(k+k')+g_{2\perp}(k+k') \bigr\} \nonumber \\
	&\times  &\Biggl[ \frac{1}{\sqrt{2}}
	\bigl( a_{k'\uparrow}^{\dagger}b_{-k'\downarrow}^{\dagger}
	+a_{k'\downarrow}^{\dagger}b_{-k'\uparrow}^{\dagger} \bigr)
	\frac{1}{\sqrt{2}}
	\bigl( b_{-k\downarrow}a_{k\uparrow} +b_{-k\uparrow}a_{k\downarrow}
	\bigr) \Biggr] \nonumber \\
	&+ &\bigl\{ g_{1\perp}(k+k')+g_{2\perp}(k+k') \bigr\} \nonumber \\
	&\times & \Biggl[ \frac{1}{\sqrt{2}}
	\bigl( a_{k'\uparrow}^{\dagger}b_{-k'\downarrow}^{\dagger}
	-a_{k'\downarrow}^{\dagger}b_{-k'\uparrow}^{\dagger} \bigr)
	\frac{1}{\sqrt{2}}
	\bigl( b_{-k\downarrow}a_{k\uparrow} -b_{-k\uparrow}a_{k\downarrow}
	\bigr) \Biggr] .\nonumber \\
	\,
	\end{eqnarray}
	The pairing interactions both for triplet $V^{\rm t}$ and singlet $V^{\rm s}$
	are thus given by
	\begin{eqnarray}
	V_{||}^{\rm t}(k+k')&=& -g_{1||}(k+k') +g_{2||}(k+k'), \label{V1dgt1}\\
	V_{\perp}^{\rm t}(k+k')&=& -g_{1\perp}(k+k') +g_{2\perp}(k+k'), \label{V1dgt2}\\
	V^{\rm s}(k+k')&=& g_{1\perp}(k+k') +g_{2\perp}(k+k') . \label{V1dgs}
	\end{eqnarray}

\section{Odd-energy superconductivity in 1d systems} \label{AppB}
	
\begin{figure}[tbp]
\begin{center}\leavevmode
\rotatebox{0}{\includegraphics[width=6cm]{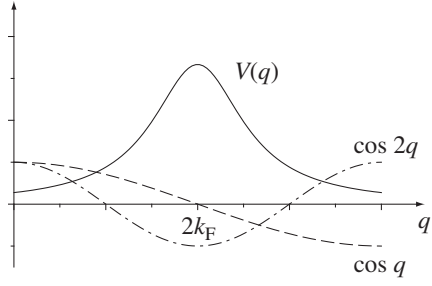}}
\caption{Illustration of the pairing interaction $V(q)$ whose 
maximum is at $q=2k_F$ (solid line).
The dashed and dash-dotted line indicates $\cos q$ and $\cos 2q$, respectively.
Here we set $2k_F = \pi /2$, assuming the quarter-filling case.
}\label{V1dcos}\end{center}\end{figure}

	In this appendix, we present a semi-analytic argument to find the most stable type of the gap symmetry.
	Despite its simplicity, it gives the gap symmetry consistent with the numerical results in q1d system.
	In order to make our argument as simple and transparent as possible,
	we here discuss the superconductivity in 1d,
	although any kind of phase transition will not be realized in purely 1d systems
	due to thermal (and quantum) fluctuations.
	We assume that the pairing interaction have a maximum at $q=2k_F $ (see Fig. \ref{V1dcos}).
	Considering that the first correction to the vertex have the $q$-dependence
	as $\log |q-2k_F |$ and the renormalized vertex is proportional to $|q-2k_F |^{\alpha}$,
	this assumption is quite natural in 1d systems even if we take into account the suppressions
	due to the temperature or the interchain hopping.

	Since the pairing interaction is an even function of momentum with period $2\pi$,
	we can expand as
	\begin{eqnarray}
	V(q) = \frac{v_0}{2} + v_1 \cos q + v_2 \cos 2q + \cdots .
	\label{Vexpand}
	\end{eqnarray}
	Here the coefficients $v_n$ are given by
	\begin{eqnarray}
	v_n = \frac{2}{\pi}\int_0^{\pi}\!\! dq \,V(q) \cos nq.
	\end{eqnarray}
	They can be written in the separable form
	\begin{eqnarray}
	V^{\rm even}(k-k')&=& \frac{v_0}{2}+v_1 \cos k \cos k' + v_2 \cos 2k \cos 2k' + \cdots , \\
	V^{\rm odd}(k-k')&=& v_1 \sin k \sin k' + v_2 \sin 2k \sin 2k' + \cdots ,
	\end{eqnarray}
	each term corresponding to the pair scattering $(k, -k) \to (k', -k')$.
	At quarter filling, the coefficients may be estimated as
	\begin{eqnarray}
	v_0 >0 ,\hspace{0.5cm} v_1 \sim 0,\hspace{0.5cm} v_2 <0
	\label{vn}
	\end{eqnarray}
	for the interaction which is peaked at $q=2k_F$.
	(The auxiliary cosine curves in Fig. \ref{V1dcos} will be helpful for a rough estimate of $v_n$.
	\cite{quarter})
%

\subsection{Spin-fluctuation model} \index{spin fluctuation model}
	If we adopt the spin-fluctuation model,
	the pairing interactions in the singlet and triplet channels are written as
	\begin{eqnarray}
	V^{\rm s} &\simeq & \frac{3v_0}{2}+3v_1 \cos k \cos k' + 3v_2 \cos 2k \cos 2k' +\cdots , 
	\label{Vcoss}\\
	V^{\rm t} &\simeq & -v_1 \sin k \sin k' -v_2 \sin 2k \sin 2k' +\cdots . \label{Vcost}
	\end{eqnarray}
	From (\ref{vn}),
	the most attractive component is $n=2$ in the spin singlet channel,
	having the gap function $\cos 2k$.
	If we consider the higher-order components,
	the pairing interaction peaked at $q=2k_F$ will favor singlet pairing with
	gap function whose general form may be written as
	\begin{eqnarray}
	\Delta^{\rm s} (k) = \sum_n \Delta_n \cos 2(2n+1)k,
	\end{eqnarray}
	which vanishes just at $k=k_F (=\pi /4)$, i.e., odd-energy gap function $\Delta (k) \propto \xi (k)$
	at quarter filling.
	This is because $v_3 \sim 0, v_4 >0, v_5 \sim 0, v_6 <0$, and so on.
	In case of half filling,
	the gap function has the form $\Delta (k) = \sum_n \Delta_n \cos (2n+1)k$,
	which also vanishes just at $k=k_F (=\pi /2 )$.

	The triplet components can also be attractive although subdominant.
	The higher-order component $v_4 >0$, however, brings moderately large attraction,
	though $v_4 < |v_2 |$.
	Therefore, the odd-energy gap function
	\begin{eqnarray}
	\Delta^{\rm t}(k) \simeq \sum_n \Delta_n \sin 4nk,
	\end{eqnarray}
	which vanishes just at $k=k_F (=\pi /4)$,
	is favored also in the triplet channel.
	We may note that as the pairing interaction becomes more singular,
	$v_4 $ becomes comparable to $v_2$,
	and the difference in $\Tc$ between singlet and triplet
	becomes smaller.
	For example, the interaction $V(q) \sim \delta (q-2k_F)$ leads to $|v_2 |= v_4$.

	We have seen that the pairing interaction whose maximum is at $q=2k_F$
	generally leads to the odd-energy gap function both for singlet and triplet pairing,
	and the corresponding coupling constant for the singlet is (three times or more)
	larger than that of triplet.
	If we consider the charge fluctuation, this difference is reduced.
	This general conclusion is completely consistent with the numerical results\cite{Fuseya}.

\subsection{1d Fermi gas model}\index{1d Fermi gas model}
	
	Next we consider the 1d Fermi gas model with repulsive interaction.
	From eqs. (\ref{V1dgt1})-(\ref{V1dgs}),
	the pairing interaction is given by
	\begin{eqnarray}
	V^{\rm s} (k+k') &=&g_1 (k+k') +g_2 (k+k'), \\
	V^{\rm t} (k+k') &=& -g_1 (k+k') +g_2 (k+k') .
	\end{eqnarray}
	We assume that both $g_1 (q)$ and $g_2 (q)$ have a maximum at $q=2k_F$,
	and expand similarly to eq. (\ref{Vexpand}) to obtain the separable form
	\begin{eqnarray}
	g_1 (k+k') &=& \frac{g_1^{(0)}}{2}+g_1^{(1)}\cos k \cos k' +g_1^{(2)} \cos 2k \cos 2k' +\cdots \\
	& &-g_1^{(1)}\sin k \sin k' -g_1^{(2)} \sin 2k \sin 2k' -\cdots , \\
	g_2 (k+k') &=& \frac{g_2^{(0)}}{2}+g_2^{(1)}\cos k \cos k' +g_2^{(2)} \cos 2k \cos 2k' +\cdots \\
	& &-g_2^{(1)}\sin k \sin k' -g_2^{(2)} \sin 2k \sin 2k' -\cdots ,
	\end{eqnarray}
	and
	\begin{eqnarray}
	V^{\rm s}(k+k') &=& \frac{g_1^{(0)}+g_2^{(0)}}{2} + (g_1^{(1)}+g_2^{(1)})\cos k \cos k' \nonumber \\
	& & \hspace{1cm}+(g_1^{(2)}+g_2^{(2)})\cos 2k \cos 2k' +\cdots , \\
	V^{\rm t}(k+k') &=& (g_1^{(1)}-g_2^{(1)})\sin k \sin k'
	+(g_1^{(2)}-g_2^{(2)})\sin 2k \sin 2k' +\cdots . \nonumber \\
	\,
	\end{eqnarray}
	In the same manner as the spin-fluctuation model,
	we can estimate each component as $g_{1, 2}^{(0)}>0$, $g_{1, 2}^{(1)}\sim +0$, 
	$g_{1, 2}^{(2)}<0$.
	Therefore, for singlet pairing, the odd-energy gap function
	is realized as in the spin-fluctuation model.
	For triplet pairing, on the other hand,
	two types of gap functions are possible.
	First, in the case that the backward scattering is dominant, i.e., $|g_1^{(n)}|>|g_2^{(n)}|$,
	the attraction originates from $n=2(2m+1)$ components $g_1^{(2)}-g_2^{(2)}<0$,
	so that the gap function should have the form
	\begin{eqnarray}
	\Delta^{\rm t}(k) =\sum_n \Delta_n \sin 2(2n+1)k ,
	\end{eqnarray}
	which has a peak at $k=k_F (=\pi /4)$, namely, $p$-wave triplet.
	Second, in the opposite case, i.e., $|g_1^{(n)}|<|g_2^{(n)}|$,
	the large attraction comes from $n=4m$, and we have
	the odd-energy gap function
	\begin{eqnarray}
	\Delta^{\rm t}(k) =\sum_n \Delta_n \sin 4n k ,
	\end{eqnarray}
	which vanishes at $k=k_F (=\pi /4)$.
	Above results are also consistent with the results of the present RG method.
	The important knowledge here is that the dominant backward scattering
	promote the $p$-wave triplet superconductivity, which competes with and can prevail over 
	the $d$-wave singlet one in 1d systems.



\begin{thebibliography}{99} 

	\bibitem{Jerome} D. Jerome, A. Mazaud, M. Ribault, K. Bechgaard: J. Physique Lett. \textbf{41}
	(1980) L95.
	
	\bibitem{TMTSFphase}
	D. J\'{e}rome: Mol. Cryst. Liq. Cryst. {\bf 79} (1982) 155.
	
	\bibitem{MSV}
	K. Miyake, S. Schmitt-Rink and C. M. Varma: Phys. Rev. B \textbf{34}
	(1986) 6554.
	
	\bibitem{Scalapino}
	D. J. Scalapino, E. Loh, Jr. and J. E. Hirsch:
	Phys. Rev. B \textbf{34} (1986) 8190.
	
	\bibitem{Emery}
	V. J. Emery: Synthetic Metals \textbf{13} (1986) 21.
	
	\bibitem{Hc2} I. J. Lee, P. M. Chaikin and M. J. Naughton:
 	Phys. Rev. Lett. {\bf 78} (1997) 3555;
 
 	\bibitem{Hc22} I. J. Lee, P. M. Chaikin and M. J. Naughton: Phys. Rev. B {\bf 62} (2000) R14669;
 	{\it ibid}. {\bf 65} (2002) 180502.
 
 	\bibitem{Knight} I. J. Lee, S. E. Brown, W. G. Clark, M. J. Strouse, M. J. Naughton,
 	W. Kang and P. M. Chaikin: Phys. Rev. Lett. {\bf 88} (2002) 017004.
 
 	\bibitem{Knight2} 
 	I. J. Lee, D. S. Chow, W. G. Clark, M. J. Strouse, M. J. Naughton,
 	P. M. Chaikin and S. E. Brouwn: Phys. Rev. B {\bf 68} (2003) 092510.
 	
 	\bibitem{Fujiwara}
	N. Fujiwara, N. M\^ori, Y. Uwatoko, T. Matsumoto, N. Motoyama and S. Uchida:
	Phys. Rev. Lett. {\bf 90} (2003) 137001.
	
	\bibitem{Shimahara} H. Shimahara: J. Phys. Soc. Jpn. {\bf 58} (1989) 1735.
	
	\bibitem{Kino} H. Kino and H. Kontani: J. Low Temp. Phys. {\bf 117} (1999) 317.
	
	\bibitem{KurokiFLEX} K. Kuroki, R. Arita and H. Aoki: Phys. Rev. B {\bf 63} (2001) 094509.
	
	\bibitem{Fuseya} Y. Fuseya, Y. Onishi, H. Kohno and K. Miyake: J. Phys.: Condens. Matter
	{\bf 14} (2002) L655.
	
	\bibitem{Tanaka} Y. Tanaka and K. Kuroki: cond-mat/0402672.
	
	\bibitem{Nomura} T. Nomura and K. Yamada: J. Phys. Soc. Jpn. {\bf 70} (2001) 2694.
	
	\bibitem{KurokiQMC} K. Kuroki, Y. Tanaka, T. Kimura and R. Arita: cond-mat/0307553.
 	
 	\bibitem{FLEXnote} If we consider the extended Hubbard model with on-site ($U$) and off-site ($V$) repulsive interactions,
   both charge and spin fluctuations can be taken into account by RPA-type approximation \cite{Tanaka};
 	neverthless, we cannot take into account the vertex correction type singularities.
	
	\bibitem{Solyom} 
	See, for example, J. S\'{o}lyom: Adv. Phys. {\bf 28} (1979) 201.
	
	\bibitem{Suzumura} Y. Suzumura and H. Fukuyama: J. Low. Temp. Phys. {\bf 31} (1978) 273.
	
	\bibitem{1loop} One-loop approximation is sufficient for $g_{1||}\ge |g_{1\perp}|$,
	 since the coupling constants are renormalized to zero\cite{Solyom}.
	 
	\bibitem{SolyomI} N. Menyh\'{a}rd and J. S\'{o}lyom: J. Low Temp. Phys. {\bf 12} (1973) 529.
	
	\bibitem{Yamaji} K. Yamaji: J. Phys. Soc. Jpn. {\bf 51} (1982) 1361.
	
	\bibitem{thesis} Detailed discussion is given in Y. Fuseya: doctral thesis, Osaka University, (2004), Chap. 5.
	
	\bibitem{note} Note that here we put $q=k+k'$.
	This is different from the standard notation $q=k-k'$, which gives
	$V^{\rm t, RPA}(q)=-U^2 \chi_0 (q)$.
	
	\bibitem{sasaki} Recently, analogous discussion was given by the third-order perturbation study
	on a model for $\beta$-Na$_{0.33}$V$_2$O$_5$, with a conclusion 
	that the $p$TS is possible due to the one-dimensional effect.
	[S. Sasaki, H. Ikeda and K. Yamada: J. Phys. Soc. Jpn., \textbf{73} (2004) 815.]
	
	\bibitem{quarter} The situation illustrated in Fig. \ref{V1dcos} is for the quarter-filling case 
	$2k_F =\pi /2$,
	but the present discussion can also be applied to the half-filling case.

\end{thebibliography}
\end{document}